\documentclass[journal,onecolumn,a4paper,12pt]{IEEEtran}
\IEEEoverridecommandlockouts

\usepackage{hyperref}
\usepackage[cmex10]{amsmath}
\usepackage{amssymb,amsfonts}
\usepackage{flushend}
\usepackage[ruled,vlined]{algorithm2e}
\usepackage{graphicx}
\graphicspath{{Figures/PDF/}{Figures/PNG/}}
\usepackage{setspace}
\usepackage{booktabs}
\usepackage{siunitx}
\usepackage[numbers,compress]{natbib}
\usepackage{texnames}
\usepackage{bm,bbm}
\usepackage{orcidlink}

\usepackage[capitalize]{cleveref}
\crefname{section}{Sec.}{Secs.}
\Crefname{section}{Section}{Sections}
\Crefname{table}{Table}{Tables}
\crefname{table}{Tab.}{Tabs.}

\begin{document}

\title{\textit{TopoFormer}: Integrating Transformers and ConvLSTMs for Coastal Topography Prediction}

\author{
{Santosh Munian$^{1}$, Oktay Karakuş$^{1}$, William Russell$^{2}$, Gwyn Nelson$^{2}$\\[1em]
$^{1}$ Cardiff University, School of Computer Science and Informatics, CF24 4AG, UK\\
$^{2}$ Wales Coastal Monitoring Centre, Alps Depot, Wenvoe, CF5 6AA, UK
}}
 
\maketitle
\begin{abstract}
	This paper presents \textit{TopoFormer}, a novel hybrid deep learning architecture that integrates transformer-based encoders with convolutional long short-term memory (ConvLSTM) layers for the precise prediction of topographic beach profiles referenced to elevation datums, with a particular focus on Mean Low Water Springs (MLWS) and Mean Low Water Neaps (MLWN). Accurate topographic estimation down to MLWS is critical for coastal management, navigation safety, and environmental monitoring. Leveraging a comprehensive dataset from the Wales Coastal Monitoring Centre (WCMC), consisting of over 2000 surveys across 36 coastal survey units, TopoFormer addresses key challenges in topographic prediction, including temporal variability and data gaps in survey measurements. The architecture uniquely combines multi-head attention mechanisms and ConvLSTM layers to capture both long-range dependencies and localized temporal patterns inherent in beach profiles data. TopoFormer's predictive performance was rigorously evaluated against state-of-the-art models, including DenseNet, 1D/2D CNNs, and LSTMs. While all models demonstrated strong performance, \textit{TopoFormer} achieved the lowest mean absolute error (MAE), as low as 2 cm, and provided superior accuracy in both in-distribution (ID) and out-of-distribution (OOD) evaluations. 
\end{abstract}

\doublespacing
\section{Introduction}

Accurate topographic measurements are essential for coastal applications such as flood risk assessment, erosion monitoring, habitat mapping, and navigation safety. Among these, Mean Low Water Spring (MLWS) levels are crucial as they define coastal boundaries, influence sediment transport, and support biodiversity studies in intertidal zones. For regions like Wales, with its intricate and dynamic coastlines, reliable MLWS data is vital for informed decision-making, especially in areas prone to erosion, flooding, and ecological changes \cite{kaur2019systematic, sarkar2016machine}. However, environmental and logistical challenges—such as adverse weather, high wave energy, and inaccessible shallow slopes—often hinder the collection of complete MLWS data, leaving gaps in surveys that can undermine topographic analysis and coastal management efforts \cite{l2017machine, lee2023accuracy}. 

To bridge these data gaps, researchers have turned to historical coastal records and advanced machine learning (ML) methods. Time series models, such as Long Short-Term Memory (LSTM) networks and emerging transformer-based architectures, have shown exceptional performance in capturing complex temporal dependencies and variability in trend prediction similar to topographic beach profiles \cite{ma2023tcln, raskoti2024exploring}. These approaches enable robust extrapolation of beach profiles, address missing data that do not reach MLWN/MLWS and improve the accuracy of topographic predictions. 
Building on this foundation, this study introduces \textit{TopoFormer}, a transformer-based architecture designed to predict topographic beach profiles referenced to MLWS and other low-water datums using surveys from 36 Welsh coastal sites provided by the Wales Coastal Monitoring Centre (WCMC - Data Platform (\url{https://www.wcmc.wales/data}). By integrating state-of-the-art techniques with diverse coastal datasets, TopoFormer provides a scalable and adaptable solution to improve coastal monitoring and management practices.

\section{Related Works}
Coastal monitoring involves observing and analysing coastal environments to understand and predict changes. Traditional methods, such as topographic surveys and tide gauge measurements, provided accurate insights but were labour-intensive and time-consuming. Modern advancements, including remote sensing, LiDAR, GIS, and auditory techniques like sonar, have revolutionized coastal monitoring. These technologies enable large-scale, real-time data collection, detailed topographic and bathymetric mapping, and the integration of diverse datasets for uncovering spatial patterns and trends
\cite{thakur2024comprehensive}.

Recent research has contributed significantly to advancing our understanding of coastal dynamics, water quality, and sediment transport, although several limitations remain. Hannides \emph{et al.} \cite{hannides2021us} assessed U.S. beach water quality monitoring programs, demonstrating their effectiveness in maintaining water safety but focusing solely on water quality without addressing sediment transport or beach morphology. Similarly, Thakur and Devi \cite{thakur2024comprehensive} reviewed advances in water quality monitoring devices, emphasizing materials and technological perspectives but neglecting sediment-changing aspects and coastal morphological processes. 

Suanez \emph{et al.} \cite{suanez2023using} provided an in-depth analysis of morphological dynamics during extreme water level events over 17 years of beach and dune monitoring. While valuable, their work lacks integration with predictive models or technologies to forecast future coastal dynamics. Banno \cite{banno2023can} contributed by utilizing long-term in-situ monitoring data to understand coastal changes, yet the study did not leverage advanced technologies like UAVs or acoustic monitoring for enhanced precision. Meng \emph{et al.} \cite{meng2024experimental} conducted experimental studies on beach profile evolution under various nourishment methods, offering critical insights but limited by their reliance on controlled experimental conditions that do not fully replicate real-world complexities. 

Pang \emph{et al.} \cite{pang2023coastal} conducted a comprehensive review of coastal erosion processes and their modelling in the context of climate change, highlighting significant factors but without integrating predictive technologies for future dynamics. Oliver \emph{et al.} \cite{oliver2024intertidal} and Cantelon \emph{et al.} \cite{cantelon2024interrelated} examined intertidal spring dynamics and coastal nutrient loading through geophysics, drone surveys, and in-situ monitoring. Although these studies advanced understanding of nutrient loading and groundwater dynamics, they offered limited insights into sediment transport processes and their management implications. Smith \emph{et al.} \cite{smith2024tidal} explored tidal pumping and intertidal springs' impact on coastal lagoon thermal variability, but their study did not extensively address nutrient loading or sediment dynamics.

The dynamic nature of coastal environments demands innovative approaches to monitoring and prediction, yet existing methodologies often fall short in addressing the interconnected challenges of data limitations, operational constraints, and holistic management. Coastal studies often focus narrowly on aspects like water quality or erosion, neglecting a holistic view of coastal dynamics. The lack of advanced predictive models and integration of technologies limits the ability to address complex challenges. Additionally, current topographic beach profile surveys are generally constrained to MLWS levels, restricting data collection to about 31.6 days annually and increasing operational risks and costs. Expanding data collection to the MLWN datum, where over 300 survey days are viable, could alleviate these issues. \textit{TopoFormer} addresses these challenges by introducing a transformer-based deep learning architecture to predict topographic beach profiles between MLWN and MLWS datums. Leveraging over 2000 historical surveys from Welsh coasts, it bridges data gaps, enhances predictive accuracy, and offers scalable solutions. This approach mitigates survey risks, reduces costs, and contributes to broader coastal monitoring and management strategies.

\section{The Proposed Methodology}
\subsection{Beach Profiles}
\begin{figure}[t!]
	\centering
	\includegraphics[width=\linewidth]{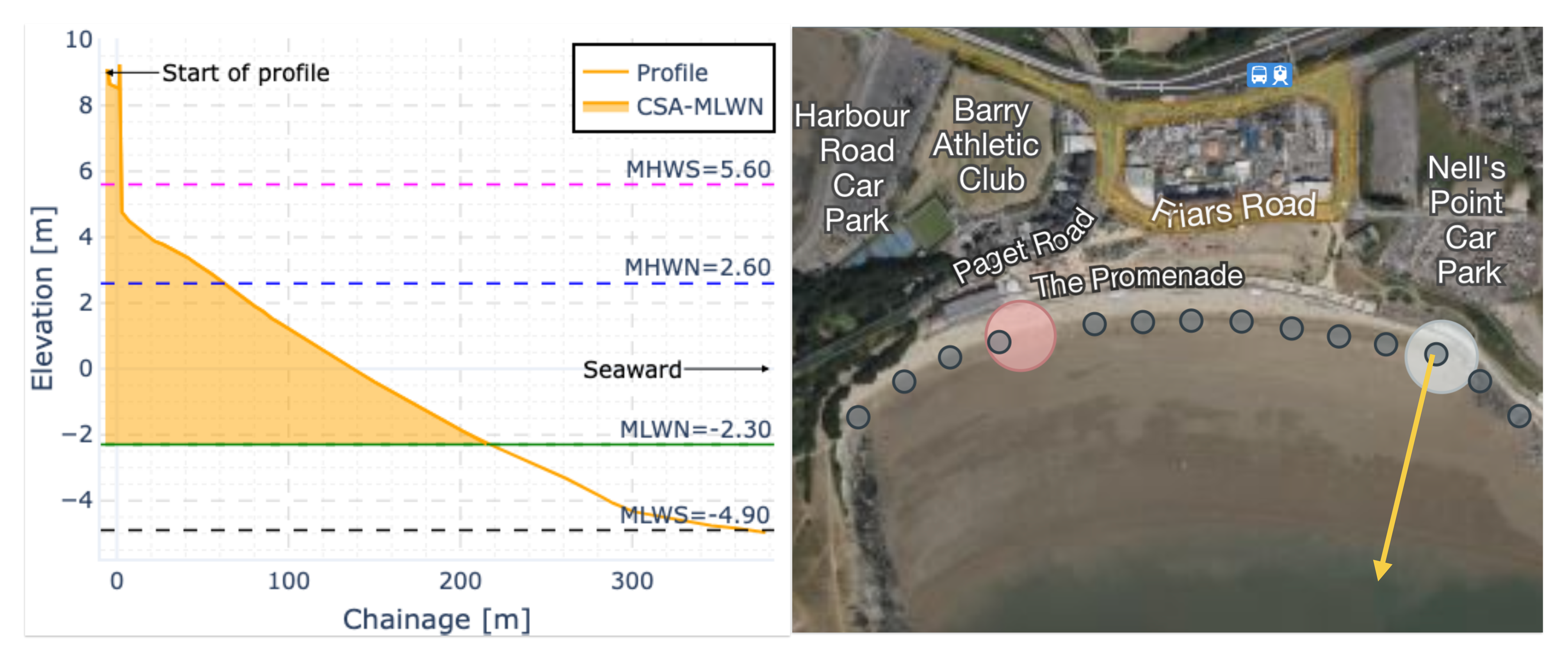}
	\caption{(LEFT) Example profile for Whitmore Bay with low and high mean tidal datums and MLWN-based CSA. (RIGHT) Satellite imagery of Whitmore Bay showing survey units and the example survey point.}\label{fig:profile}\vspace{-0.25cm}
\end{figure}

Beach profiles are 2D measurements of elevation and distance (chainage) along a pre-determined line. It is measured from the back of the beach such as a sea wall and extend seaward until the MLWS elevation is reached (this is known as a vertical datum). An example profile and its satellite imagery representation are shown in Figure \ref{fig:profile}. Over time beach profiles are repeated and their changes can be compared in a graph. The four vertical datums and the 'back of beach' chainage are used to calculate cross sectional area (see Figure \ref{fig:profile}).

The four vertical datums are: (1-2) MHWS (Mean High Water Spring) and MHWN (Mean High Water Neap) are the average heights of high waters of the spring and neap tides, respectively. (3-4) MLWN and MLWS are, on the other hand, the average height of low waters of the neap and spring tides, respectively. Cross Sectional Area (CSA) is the area underneath the profile line from the back of the mobile beach to a specific vertical datum e.g. MLWN. All four datums and CSA examples are shown in Figure \ref{fig:profile}.

\subsection{TopoFormer}
TopoFormer is a hybrid LSTM-Transformer-based architecture developed to accurately predict topographic beach profiles referenced to mean-water datums by combining the strengths of transformers and ConvLSTM layers, TopoFormer captures both long-range dependencies and localized temporal trends, making it a novel approach in time-series prediction for tidal applications. The model processes beach profile data in the form of sequential elevation-chainage pairs. These pairs are normalised to ensure compatibility across diverse datasets, enabling TopoFormer to generalize across multiple coastal locations and adapt to varied tidal characteristics.  

The model begins with six layers of transformer blocks, each designed to capture long-range dependencies in the profiles. Each transformer block consists of the following: (i) \textit{Multi-Head Attention:} dynamically assigns weights to different parts of the input sequence, prioritizing critical regions such as those near MLWN and MLWS. (ii) \textit{Layer Normalization:} ensures stable gradient flow and faster convergence during training. (iii) \textit{ConvLSTM Layers:} four ConvLSTM layers within each transformer block capture localized temporal and spatial dependencies, which is particularly important for understanding subtle variations in tidal profiles. 

After the transformer blocks, the output is passed to a multi-layer perceptron (MLP) consisting of two fully connected layers, designed to refine the predictions further. Finally, a dense layer generates the predicted elevation values corresponding to the chainage points, ensuring that the output aligns with the expected trends. The model is trained using a Mean Absolute Error (MAE) loss function, and the \textit{Adam} optimizer is employed with an initial learning rate of \(1 \times 10^{-3}\). Training stability is maintained through early stopping, and the monitoring is done using the validation loss.  

TopoFormer introduces key innovations that set it apart from traditional and deep learning-based topographic prediction methods: (i) it combines ConvLSTM layers with transformer blocks, effectively capturing both long-range dependencies and localized temporal trends; (ii) this integration enables simultaneous handling of spatial and temporal variations, crucial for this prediction tasks; (iii) its attention mechanism focuses on critical regions like MLWN and MLWS, ensuring high precision where it matters most; (iv) with only 761K trainable parameters, TopoFormer is computationally efficient and scalable; and (v) its modular design supports integration with additional data modalities, such as satellite imagery and meteorological data, for comprehensive coastal monitoring. 
\begin{figure}[t]
	\centering
	\includegraphics[width=0.5\linewidth]{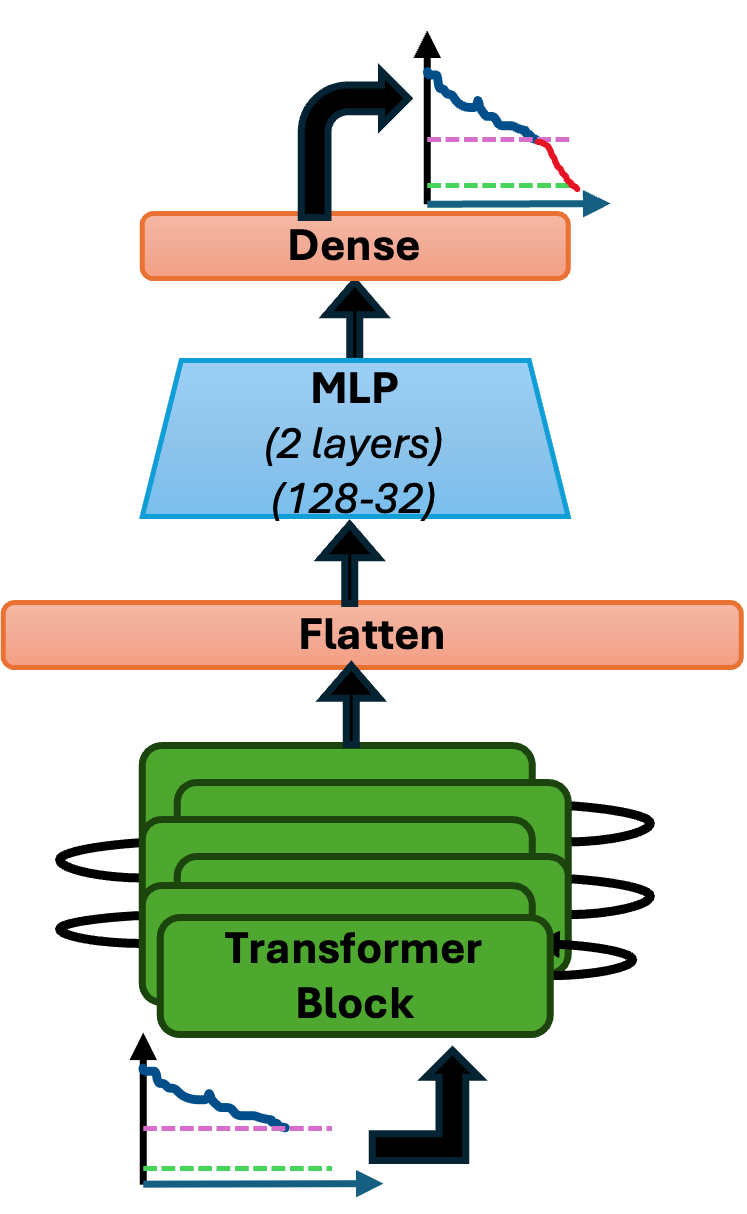}
    \includegraphics[width=0.4\linewidth]{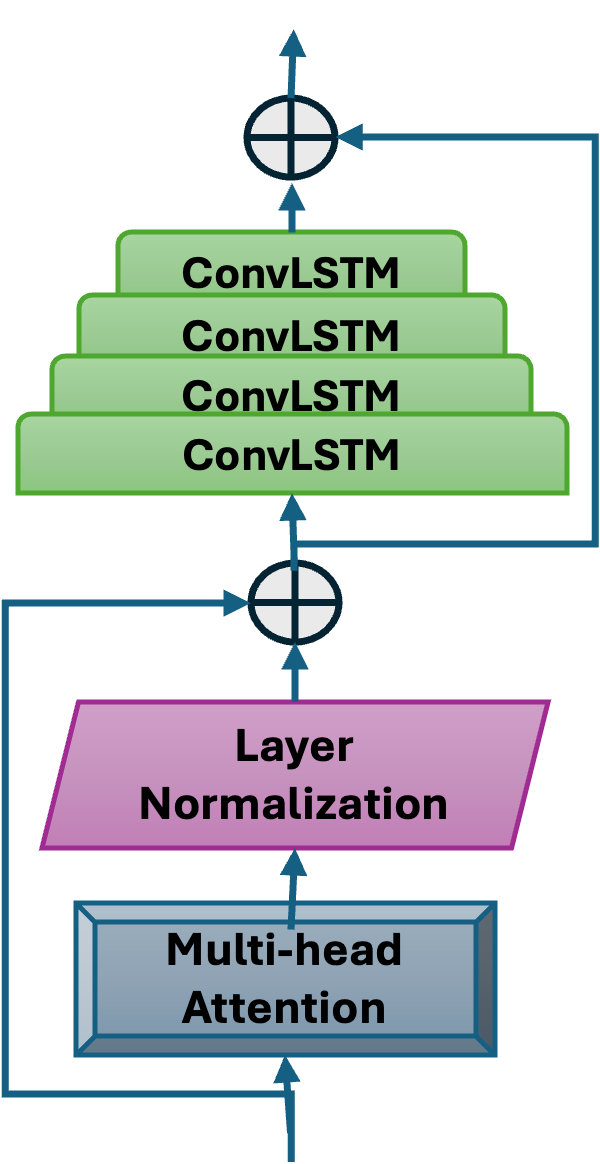}
	\caption{(LEFT) TopoFormer, and (RIGHT) Transformer block details.}\label{fig:topoFormer}
\end{figure}

Figure \ref{fig:topoFormer}-(LEFT) illustrates the sequential architecture of TopoFormer, showing its transformer blocks, MLP layers, and the final dense layer. Figure \ref{fig:topoFormer}-(RIGHT) provides an expanded view of the individual transformer block, highlighting the multi-head attention mechanism, ConvLSTM layers, and layer normalization. 

\section{Experimental Analysis}
\subsection{Dataset}
The dataset used in this study consisted of 2,757 profile measurements provided by the WCMC. After preprocessing to address missing measurements, the total number of usable profiles was reduced to 1,366. These profiles were selected based on their ability to reach MLWS across 33 different coastal locations in Wales. The dataset was divided into training, validation, and testing subsets in a 7:2:1 ratio to ensure consistent evaluation across all models. To standardize the profiles for model processing, each profile was resampled into 100 elevation-chainage pairs and was represented as an array of size (180,), where the first 80 elevation samples (extending down to MLWN) and all 100 chainage samples were included as input features. The remaining 20 elevation samples (below MLWN to MLWS) served as prediction targets. To account for varying geographical profiles, chainage values were adjusted such that the 0m elevation point corresponded to a chainage value of 0m. Negative chainage values represented points at higher elevations above sea level, while positive chainage values corresponded to points below sea level. This preprocessing ensured a consistent spatial alignment across all profiles, facilitating robust model training and evaluation.  

\subsection{Experiments \& Results}
The experimental analysis evaluated the performance of \textit{TopoFormer} against several baseline models, including DenseNet, LSTM, bi-LSTM, ConvLSTM, and 1D/2D-CNNs. The comparison was conducted using test data and focused on key performance metrics: MAE, Root Mean Square Error (RMSE), Mean Absolute Percentage Error (MAPE), and the number of trainable parameters. To further assess the robustness of TopoFormer, the analysis was extended to include predictions on out-of-distribution (OOD) data. Specifically, TopoFormer and the two best-performing baseline models were tested on profiles that did not reach MLWS for a selected survey site. This final evaluation highlighted the models' ability to generalize to incomplete or unseen data, a critical aspect in real-world coastal monitoring applications.

\begin{table}[t]
  \centering
  \caption{Model performance comparison in terms of MAE, RMSE and Number of Trainable Parameters. (\textbf{Bold} and \underline{underlined} values refer to the best and 2$^\text{nd}$ best models, respectively.)}
    \begin{tabular}{p{4cm}p{3cm}p{3cm}p{4cm}}
    \toprule
    \textbf{Model} & \textbf{MAE} & \textbf{RMSE} & \textbf{Trainable Params} \\
    \toprule
    \textit{DenseNet} & 0.133 & 0.164 & \underline{253K} \\
    \textit{LSTM}  & \underline{0.031} & \underline{0.038} & 387K \\
    \textit{biLSTM} & 0.062 & 0.070 & \textbf{186K} \\
    \textit{ConvLSTM} & 0.034 & 0.038 & 984K \\
    \textit{1D-CNN} & 0.054 & 0.064 & 2,981K \\
    \textit{2D-CNN} & 0.040 & 0.048 & 799K \\\hline
    \textbf{TopoFormer} & \textbf{0.021} & \textbf{0.026} & 761K \\
    \bottomrule
    \end{tabular}%
  \label{tab:results}%
\end{table}%

The performance comparison, as summarized in Table \ref{tab:results}, demonstrates the clear advantages of the proposed TopoFormer model in terms of prediction accuracy and computational efficiency. Among all the models evaluated, TopoFormer achieved the lowest values for both MAE (0.021) and RMSE (0.026), indicating its superior ability to predict topographic elevations with high precision. While models such as LSTM and ConvLSTM performed well, with MAE and RMSE values of 0.031 and 0.034, respectively, they fell short of matching the accuracy achieved by TopoFormer. Notably, the biLSTM model had the smallest number of trainable parameters (186K), but its MAE and RMSE metrics (0.062 and 0.070) were significantly less competitive, illustrating a trade-off between model complexity and prediction accuracy. Furthermore, TopoFormer maintained a balanced architecture with 761K trainable parameters, which is competitive when compared to other models. For example, ConvLSTM, while comparable in error metrics, was substantially more computationally intensive with 984K parameters. Although LSTM achieved the second-best MAE (0.031) and RMSE (0.038), it required fewer parameters (387K) compared to TopoFormer (761K). Models like 1D-CNN and 2D-CNN, with trainable parameters reaching 2,981K and 799K, respectively, highlighted the significant computational demand of convolutional architectures without delivering better accuracy.

\begin{figure}[t]
	\centering
	\includegraphics[width=0.8\linewidth]{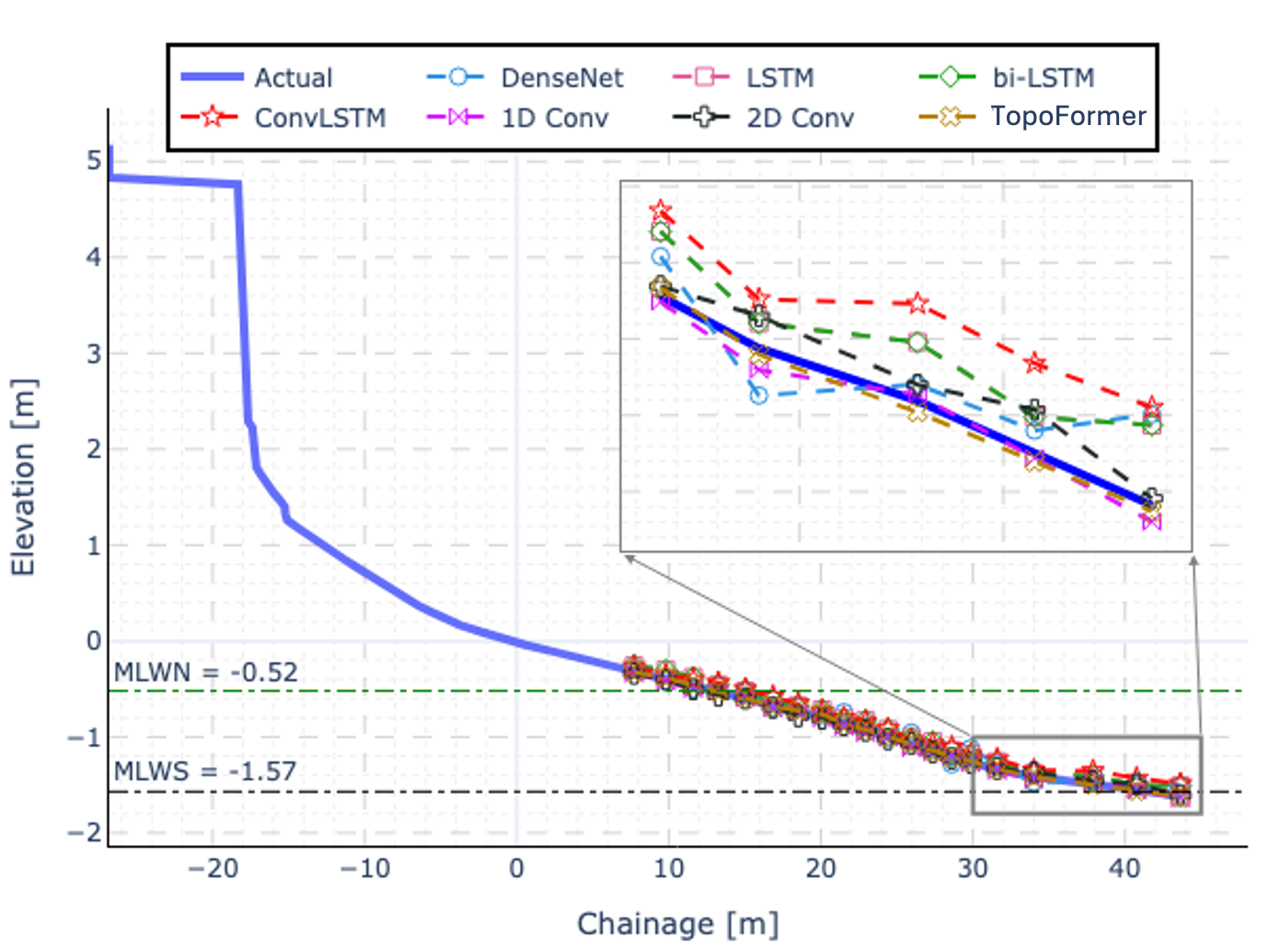}
	\caption{Elevation predictions for MLWN-MLWS range for all models.}\label{fig:predictMLWS}
\end{figure}

\begin{figure}[t]
	\centering
	\includegraphics[width=0.8\linewidth]{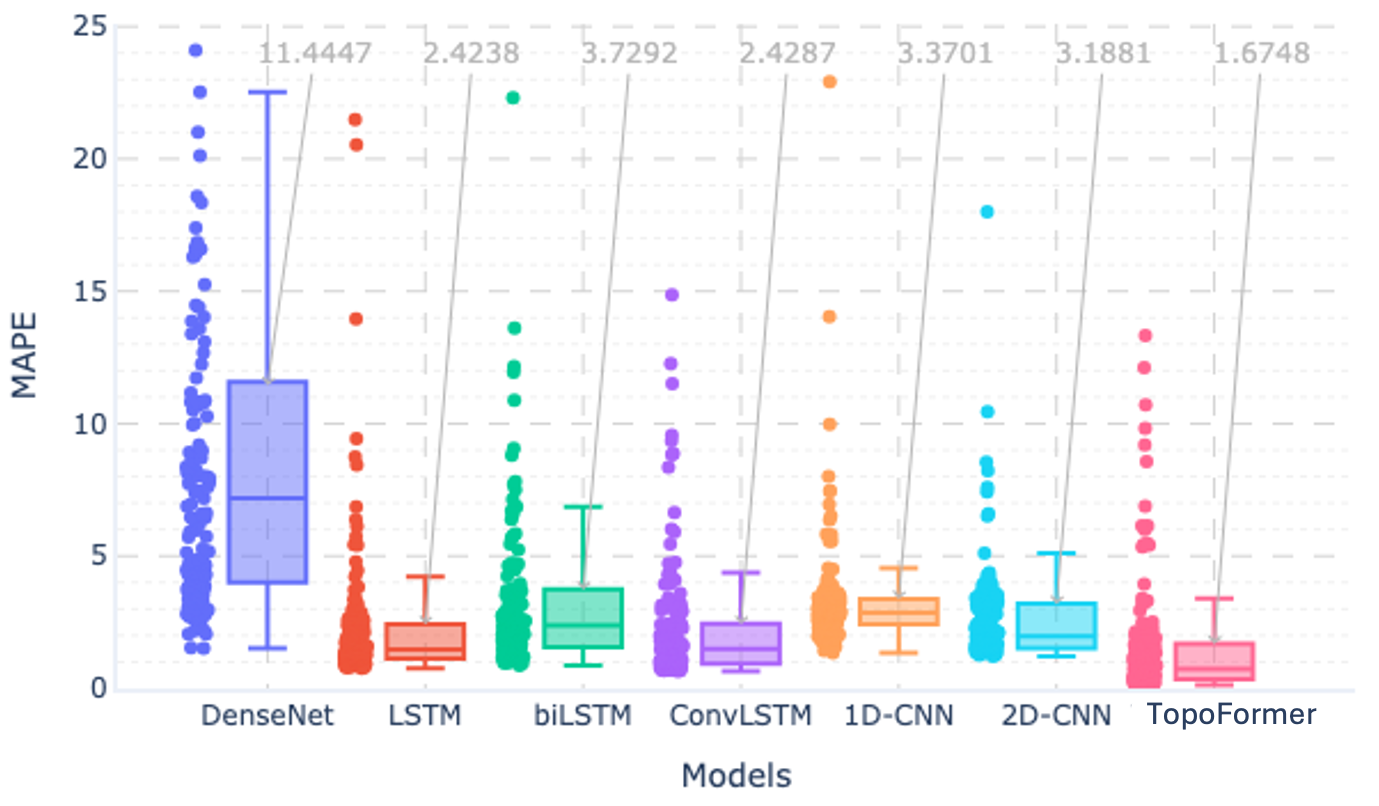}
	\caption{MAPE box-plots for test set with 75\% quantiles highlighted.}\label{fig:MAPE}
\end{figure}

Figure \ref{fig:predictMLWS} compares the prediction performance of all models for an example profile. While this visual analysis indicates that all models perform similarly, the zoomed-in section reveals TopoFormer's accuracy in predicting the range between MLWN and MLWS. This highlights TopoFormer's capability to capture subtle but critical variations in beach profiles, making it particularly effective for accurate topographical mean low-water predictions. Figure \ref{fig:MAPE} provides a box plot of the MAPE for all models, offering descriptive statistics to evaluate their performance distributions. TopoFormer exhibits the lowest MAPE across all statistical metrics, with a notably low 75th percentile value of 1.6748, compared to LSTM's 2.4238 and ConvLSTM's 2.4287. Furthermore, TopoFormer achieves the smallest minimum and 25th percentile values, demonstrating its consistent accuracy and robustness across the dataset.

The results strongly advocate for TopoFormer's utility in topographic prediction tasks, as it combines exceptional accuracy with a manageable number of parameters. Its ability to outperform traditional models and even deep-learning-based baselines validates the effectiveness of its transformer-based design for this application. By addressing the shortcomings of existing approaches, TopoFormer represents a meaningful advancement in topographical prediction and offers a robust solution for real-world coastal monitoring challenges.

\begin{figure}[t]
	\centering
	\includegraphics[width=0.8\linewidth]{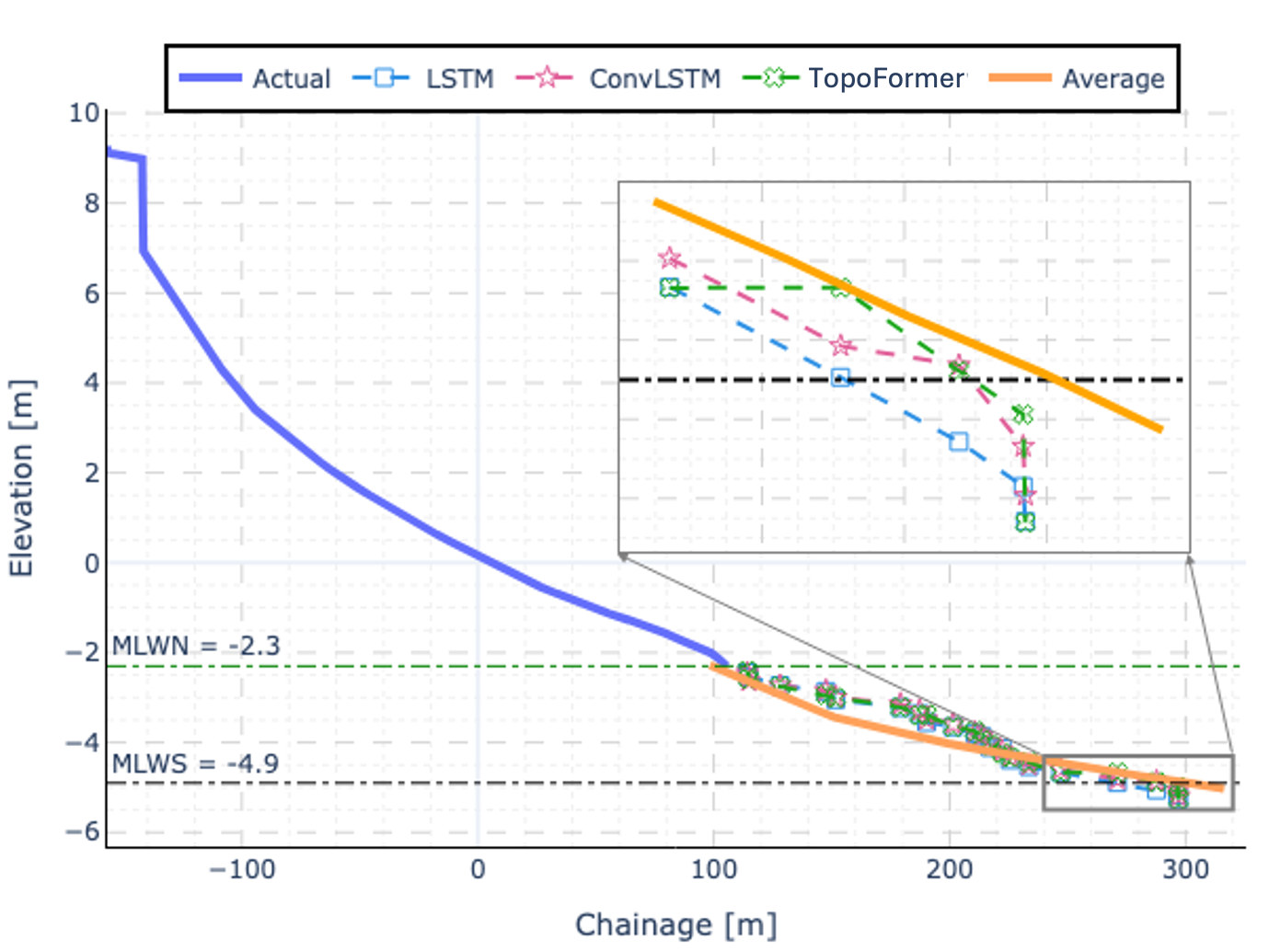}
	\caption{OOD elevation predictions down to MLWS with comparison to average profile from actual profile measurements.}\label{fig:predictMLWN}
\end{figure}

For the final analysis, an out-of-distribution (OOD) profile with measurements reaching only MLWN was used to evaluate the predictive performance of the three best-performing models: TopoFormer, LSTM, and ConvLSTM. The objective was to predict elevations down to MLWS for this profile. Since no ground truth exists for this specific date, an average profile derived from existing measurements between MLWN and MLWS was used as a reference for comparison. As illustrated in Figure \ref{fig:predictMLWN}, all three models demonstrated strong performance in predicting the OOD profile. However, TopoFormer consistently provided predictions that closely aligned with both the average trend and the elevation-chainage pair of MLWS. This highlights TopoFormer's superior generalization capability, as it is able to effectively capture coastal dynamics even for profiles outside the training distribution. 

\section{Conclusions}
This study introduced \textit{TopoFormer}, a transformer-based deep learning model for topographic prediction tasks, demonstrating its ability to outperform traditional and state-of-the-art models on various metrics, including MAE, RMSE, and MAPE. Its transformer-based architecture proved effective in predicting both in-distribution and out-of-distribution profiles, showcasing its robustness and adaptability. By addressing key challenges in topographic prediction, TopoFormer offers a practical solution for coastal monitoring and management.  

For future work, combining satellite imagery with survey data within a multi-modal AI framework could further enhance predictive accuracy and provide richer insights into coastal dynamics. Additionally, the geo-informed development of advanced machine learning techniques tailored to topographic prediction could unlock even greater potential for addressing complex coastal challenges. 

\bibliographystyle{IEEEtranN}
\bibliography{references}

\end{document}